\documentclass[twocolumn,showpacs,preprintnumbers,amsmath,amssymb]{revtex4}
\usepackage{amsmath}

\usepackage{bm}
\usepackage[colorlinks=true,linkcolor=blue,citecolor=blue,urlcolor=blue]{hyperref}
\usepackage{color}

\usepackage{graphicx}
\usepackage{epstopdf}

\begin{document}


\title{Size dependent magnetic and electrical properties of Ba-doped nanocrystalline BiFeO$_3$ }


\author{Mehedi Hasan}
\email[Author to whom correspondence should be addressed (e-mail): ]{mhrizvi@gce.buet.ac.bd}
\affiliation{Department of Glass and Ceramic Engineering, Bangladesh University of Engineering and Technology, Dhaka 1000, Bangladesh}

\author{M.A. Hakim}
\affiliation{Department of Glass and Ceramic Engineering, Bangladesh University of Engineering and Technology, Dhaka 1000, Bangladesh.}

\author{M. A. Basith}
\email[Author to whom correspondence should be addressed (e-mail): ]{mabasith@phy.buet.ac.bd}
\affiliation{Department of Physics, Bangladesh University of Engineering and Technology, Dhaka-1000, Bangladesh.
}

\author{Md. Sarowar Hossain }
\affiliation{S. N. Bose National Centre for Basic Sciences, Salt Lake City, Kolkata, West Bengal 700098, India.
}
\author{Bashir Ahmmad}
\affiliation{Graduate School of Science and Engineering, Yamagata University, 4-3-16 Jonan, Yonezawa 992-8510, Japan.}

\author{M. A. Zubair, A. Hussain and Md. Fakhrul Islam }
\affiliation{Department of Glass and Ceramic Engineering, Bangladesh University of Engineering and Technology, Dhaka, Bangladesh.}
\date{\today}

\begin{abstract}
	Improvement in magnetic and electrical properties of multiferroic BiFeO$_3$ in conjunction with their dependence on particle size is crucial due to its potential applications in multifunctional miniaturized devices. In this investigation, we report a study on particle size dependent structural, magnetic and electrical properties of sol-gel derived Bi$_{0.9}$Ba$_{0.1}$FeO$_3$ nanoparticles of different sizes ranging from $\sim$ 12 to 49 nm. The substitution of Bi by Ba significantly suppresses oxygen vacancies, reduces leakage current density and Fe$^{2+}$ state. An improvement in both magnetic and electrical properties is observed for 10 \% Ba-doped BiFeO$_3$ nanoparticles compared to its undoped counterpart. The saturation magnetization of Bi$_{0.9}$Ba$_{0.1}$FeO$_3$ nanoparticles increase with reducing particle size in contrast with a decreasing trend of ferroelectric polarization. Moreover, a first order metamagnetic transition is noticed for $\sim$ 49 nm Bi$_{0.9}$Ba$_{0.1}$FeO$_3$ nanoparticles which disappeared with decreasing particle size. The observed strong size dependent multiferroic properties are attributed to the complex interaction between vacancy induced crystallographic defects, multiple valence states of Fe, uncompensated surface spins, crystallographic distortion and suppression of spiral spin cycloid of BiFeO$_3$.

\end{abstract}

\maketitle
\section{Introduction} \label{I}
The coexistence of ferroelectricity and magnetism in a single phase is a quest for many technological applications including sensors, magnetic recording media and spintronic devices \cite{ref1,ref2,ref3}. Although ferroelectricity and magnetism tend to be mutually exclusive their coexistence can only be evidenced in rare materials called multiferroics \cite{ref1}.  Recently, there is enormous interest in the study of multiferroic BiFeO$_3$ (BFO) ceramic materials because of its promising applications in the fundamental research and device applications owing to its high ferroelectric Curie temperature ( T$_c$ $\sim$ 1103K) and antiferromagnetic N\'eel temperature \textcolor {red} {(T$_N$ $\sim$ 643K)} \cite{ref2,ref3,ref4}. Many of these investigations are focused on improving magnetic and ferroelectric properties of BFO. Compared to magnetic properties, significant improvements in electrical properties have been achieved in bulk single crystal \cite{ref201} and strained thin films \cite{ref5,ref6}. The low magnetic moment and degraded ferroelectric properties due to high leakage current density are the mostly reported limitations for the potential applications of multiferroic BFO \cite{ref7}. The high leakage current density in undoped BFO is attributed to the highly volatile nature of Bi with its corresponding off-stoichiometric phases and oxygen vacancies \cite{ref8}. On the other hand, BFO possesses G-type antiferromagnetic spiral modulated spin structure (SMSS) with a period of ̴ 62 nm which continues through the crystal and thereby cancel out macroscopic magnetization in bulk \textcolor {red} {\cite{ref202}}. Previous investigations have shown that doping with rare-earth, alkaline-earth and transition elements modify spiral spin structure and also reduce leakage current which results in improved magnetic as well as ferroelectric properties \cite{ref7,ref12,ref13}. Besides, recent investigations have demonstrated enhanced magnetization in BFO nanostructures with sizes less than 62 nm by virtue of its modified spiral spin structure \cite{ref9,ref10,ref11}. 

The effect of doping concentration on multiferroic properties of both bulk and nanocrystalline Ba-doped BFO ceramic materials were reported in Refs. \cite{ref14,ref15,ref16}. Particularly, in Ref. \cite{ref15} Ba doped Bi$_{1-x}$Ba$_{x}$FeO$_3$ (x = 0.05-0.30) nanocrystallines were prepared by a sol-gel method to investigate the effect of Ba doping on magnetic and dielectric properties. The average grain size of the prepared samples was found to vary from 20 to 70 nm by varying the composition of that alloy system. In our investigation, we have prepared 10 \% Ba-doped BFO nanoparticles to explore the simultaneous effect of cation substitution and size confinement on their multiferroic properties. Notably, to inspect the intrinsic size effect, here the particle sizes were varied thermodynamically for a fixed doping concentration of Ba. The concentration of Ba-doping has been chosen 10 \% due to its optimum magnetic and electrical properties \cite{ref15,ref16,ref17}. We have delineated the preparation process of Bi$_{0.9}$Ba$_{0.1}$FeO$_3$ (BBFO) nanoparticles to report the effect of thermodynamically varied  particle size on crystal structure, cation valence sate, oxygen vacancy and corresponding magnetic, ferroelectric and leakage behavior. The magnetic and electrical properties were found to be enhanced through combined effect of cation substitution and size confinement in Ba-doped BFO nanoparticles.

 
The presence of metamagnetic transition is another intriguing feature observed in this Ba-doped nanoparticles system.  The metamagnetic transition is referred to a jump of initial magnetization curve to a higher value with increasing magnetic field. In case of perovskite oxides the coexistence of competing ferromagnetic and antiferromagnetic phases plays a critical role for metamagnetic transition to take place \cite{ref203,ref204}. It is worth mentioning that in the present investigation we have also observed a size dependent metamagnetic transition for BBFO nanoparticles which is rare to the best of our knowledge.

\section{Experimental details} \label{II}
In this investigation we utilized analytical grade pure bismuth nitrate pentahydrate (Bi(NO$_3$)$_3$.5H$_2$O), iron nitrate nonahydrate (Fe(NO$_3$)$_3$.9H$_2$O), barium nitrate (Ba(NO$_3$)$_2$), citric acid (C$_6$H$_8$O$_7$), ethylene glycol (C$_2$H$_6$O$_2$) to preapare undoped and 10\% Ba-doped BFO nanoparticles. For a typical BBFO powder synthesis process stoichiometric proportion of  Bi(NO$_3$)$_3$.5H$_2$O (0.016 mol), Fe(NO$_3$)$_3$.9H$_2$O (0.02 mol), Ba(NO$_3$)$_2$ (0.004 ml) and C$_6$H$_8$O$_7$ (0.04 mol) were dissolved in 400 ml deionized water. Subsequently the solution was heated under continuous stirring at 75-95 $^o$C for 4h to obtain precursor xerogel. The detailed synthesis procedure for the preparation of BFO precursor xerogel is described in details in our previous investigation \cite{ref11}. The ground precursor xerogel powders were annealed at 400$^o$C-600$^o$C for two hours to obtain BFO and BBFO nanoparticles. X-ray diffraction (XRD) analysis, X-ray photoelectron spectroscopy (XPS) analysis, field emission scanning electron microscope (FESEM) imaging and magnetic measurements were performed with annealed powders. Furthermore, to measure electrical properties pellets were prepared by mixing precursor xerogel powders with PVA binder followed by pressing and annealing at 400$^o$C-600$^o$C with high heating rate (20$^o$C/min) \cite{ref22}.

The crystal structure of the synthesized BFO and BBFO nanocrystals were analyzed by powder XRD technique using CuK${\alpha}$ radiation in the scanning range of 10$^o$ to 70$^o$ (model 3040–X’Pert PRO, Philips). We adopted XPS (model 1600, ULVAC–PHI Inc.) analysis to examine the presence of oxygen vacancies and oxidation states of cations in the synthesized samples. FESEM (model JSM 7600, Jeol) was used to observe particle size and morphology of the synthesized nanoparticles. During FESEM imaging the accelerating voltage was maintained at 5 kV to eliminate charging effect. To explore size dependent magnetic properties, room temperature magnetic hysteresis (M-H) measurements were carried out using a vibrating sample magnetometer (model VSM 7407, Lake Shore) up to applied field of  $\pm$16.5 kOe. To investigate leakage current density and ferroelectric polarization of prepared samples a ferroelectric loop tracer in conjunction with external amplifier (10 kV) was used (model Precision Multiferroic and Ferroelectric Test System, Radiant). 


\begin{figure}[hh]
	\centering
	\includegraphics[width=8cm]{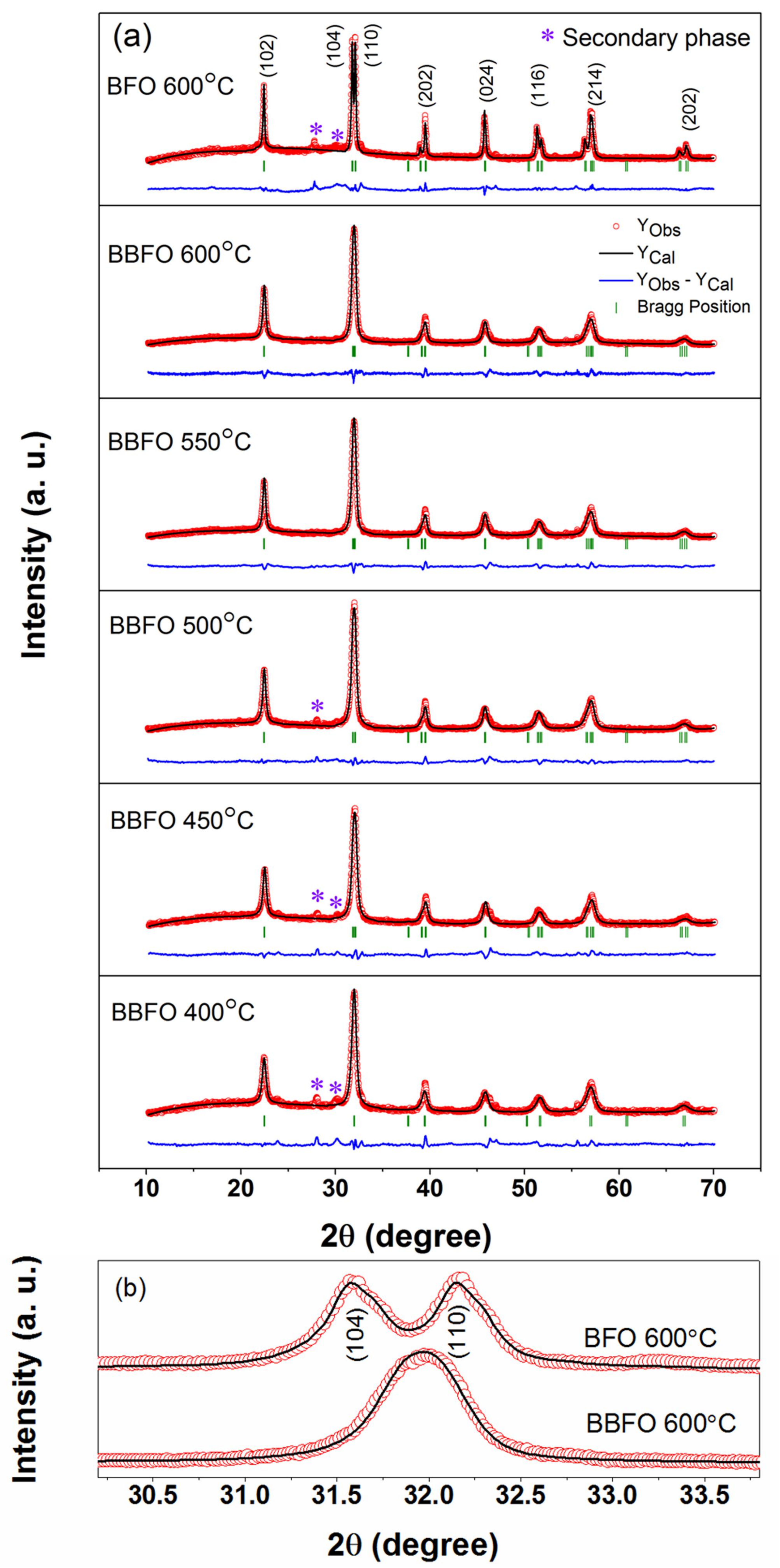}
	\caption{(a) X–ray diffraction patterns of BFO and BBFO powders annealed at different temperatures. (b) Magnified XRD patterns of (104) and (110) diffraction peaks.} \label{fig1}
\end{figure}

\section{Results and discussions} \label{III}
\subsection{Structural characterization} \label{I}
Room temperature XRD analysis and their Rietveld refinements were performed for synthesized nanoparticles to estimate the changes in crystalline parameters as a result of Ba-doping and changing their particle sizes. The observed, calculated and their difference of XRD profiles for BFO powders annealed at 600$^o$C and BBFO powders annealed at 400$^o$C-600$^o$C are shown in Fig. \ref{fig1}(a). The presence of diffraction peaks such as (102), (110), (202) and (024) confirms the formation of crystalline BBFO powders. In addition to the desired spectra some extra peaks [marked by  asterisk (‘*’) in Fig. \ref{fig1}(a)] become visible in the case of undoped BFO nanoparticles which are associated with Bi$_2$Fe$_4$O$_9$ and Bi$_{25}$FeO$_{39}$ impurity phases \cite{ref23,ref24}. A notable suppression in impurity peaks for Ba-doped nanoparticles annealed at 600$^o$C and 550$^o$C is discerned in Fig.  \ref{fig1}(a) which, however, reappeared with lowering annealing temperature. Our previous study \cite{ref11} confirmed that low temperature stability of impurity phases \cite{ref23,ref24} makes their existence favorable with lowering annealing temperature. Moreover, Fig. \ref{fig1} (b) demonstrates that the splitting of 104 and 110 peaks are depressed in case of BBFO compounds compared to its pure counterpart which has been described in previously published reports either as distortion in the rhombohedral symmetry \cite{ref25} or transition from rhombohedral to other symmetry (pseudo cubic \cite{ref16,ref26}), tetragonal \cite{ref15} and orthorhombic \cite{ref14}). Whatever the case may be, a degree of distortion in BBFO perovskite structure is expected according to the Goldschimidt tolerance factor (t) as follows:
\begin{align}
\mathrm{t = \frac{((1-x)r_{Bi}+xr_{Ba})+r_0}{\sqrt{2}(r_{Fe}+r_0)}}
\end{align}

where r$_{Bi}$, r$_{Ba}$, r$_{Fe}$ and r$_O$ are the effective ionic radii of Bi, Ba, Fe and O ions respectively. The tolerance factor for ideal cubic perovskite structure is exactly 1 and that of pure BFO is about 0.89 using Shannon ionic radii \cite{ref26,ref27}. However, it is clear from equation (1) that substitution of Bi$^{3+}$ (1.03 ${\buildrel _{\circ} \over {\mathrm{A}}}$) ion by relatively big Ba$^{2+}$ (1.35 ${\buildrel _{\circ} \over {\mathrm{A}}}$) ion increases the value of tolerance factor. The calculated tolerance factor for 10\% Ba doped BFO is approximately 0.91 which is considerably high compared to its pure counterpart. The theoretical increase in tolerance factor ratifies the modification in crystal structure as a result of Ba-doping.

\begin{table}[!h]
	\caption{Particle sizes, refined structural parameters and normalized lattice parameters of the synthesized nanoparticles.} \label{Tab1} 
	\begin{center}
		\begin{tabular}{|l|l|l|l|l|l|l|}
			\hline
			${Para}$&\multicolumn{6}{c|}{Annealing temperatures in $^o$C }  \\
			${meters}$& \multicolumn{6}{c|}{} \\
			\cline{2-7}
			&BFO &BBFO  &BBFO&BBFO&BBFO&BBFO\\
			&600 $^o$C&600 $^o$C&550 $^o$C&500 $^o$C&450 $^o$C&400 $^o$C  \\
			\hline
			Particle &86 &49&36&23&15&12\\
			size (nm) & &&&&&\\
			\hline	
			R$_{p}$ &8.26 &4.41&4.75& 4.80&4.83&5.02\\
			\hline
			R$_{wp}$ &9.89&5.48&5.98&6.34&6.37&7.47\\
			\hline
			${\chi^{2}}$ &5.67&2.64&3.32&3.93&3.78&3.87\\
			\hline
			Space &R3c &R3c&R3c&R3c&R3c&R3c\\
			group & &&&&&\\
			\hline
			a$_{hex}$(${\buildrel _{\circ} \over {\mathrm{A}}}$) &5.5747&5.5806&5.5837&5.5859&5.5901&5.5980\\
			\hline
			c$_{hex}$(${\buildrel _{\circ} \over {\mathrm{A}}}$) &13.861&13.831&13.823&13.811&13.791&13.753\\
			\hline
			V(${\buildrel _{\circ} \over {\mathrm{A}}})^3$ &373.202&373.225&373.241&373.237&373.271&373.329\\
			\hline
			a$_{n}$(${\buildrel _{\circ} \over {\mathrm{A}}}$) &3.9419&3.9461&3.9483&3.9498&3.9528&3.9584\\
			\hline
			c$_{n}$(${\buildrel _{\circ} \over {\mathrm{A}}}$) &4.0015 &3.9929&3.9906&3.9870&3.9812&3.9702\\
			\hline
				c$_{n}$/a$_{n}$ &1.0151 &1.0119&1.0107&1.0094&1.0072&1.0030\\
				\hline
			Fe-O-Fe &153.904 &161.182&161.153&159.4715&157.196&152.508\\
			${\phi^o}$ &&&&&&\\
			
			\hline
		\end{tabular}
	\end{center}
\end{table}

In addition to the effect of Ba-doping, the reduction in particle size may also modify the crystalline parameters of BBFO nanostructures. Therefore, the calculated particle sizes, refined lattice parameters along with normalized lattice parameters (a$_n$=a$_{hex}$/$\sqrt{2}$ and c$_n$=a$_{hex}$/$\sqrt{12}$)\cite{ref28} and Fe-O-Fe bond angles are enlisted in Table \ref{Tab1}. Full width at half maximum intensity (FWHM) of the (1 0 2) diffraction peak (sharp and separate) was utilized in Debye-Scherrer equation to calculate average particle sizes of the synthesized nanocrystals. The calculated average particle sizes of BBFO nanoparticles annealed at 400, 450, 500, 550 and 600$^o$C are reported to be 12 nm, 15 nm, 23 nm, 35 nm and 49 nm respectively. It is observed that normalized c/a ratio decreases with reducing particle size which has been described in previous investigation as the diminution of rhombohedral symmetry and onset of pseudo cubic symmetry \cite{ref28}. Moreover, the observed change in Fe-O-Fe angle may cause changes in the tilting of FeO$_6$ octahedron, and subsequently a modification in magnetic and electrical properties is expected for BBFO nanoparticles. 
\begin{figure}[hh]
	\centering
	\includegraphics[width=8.5cm]{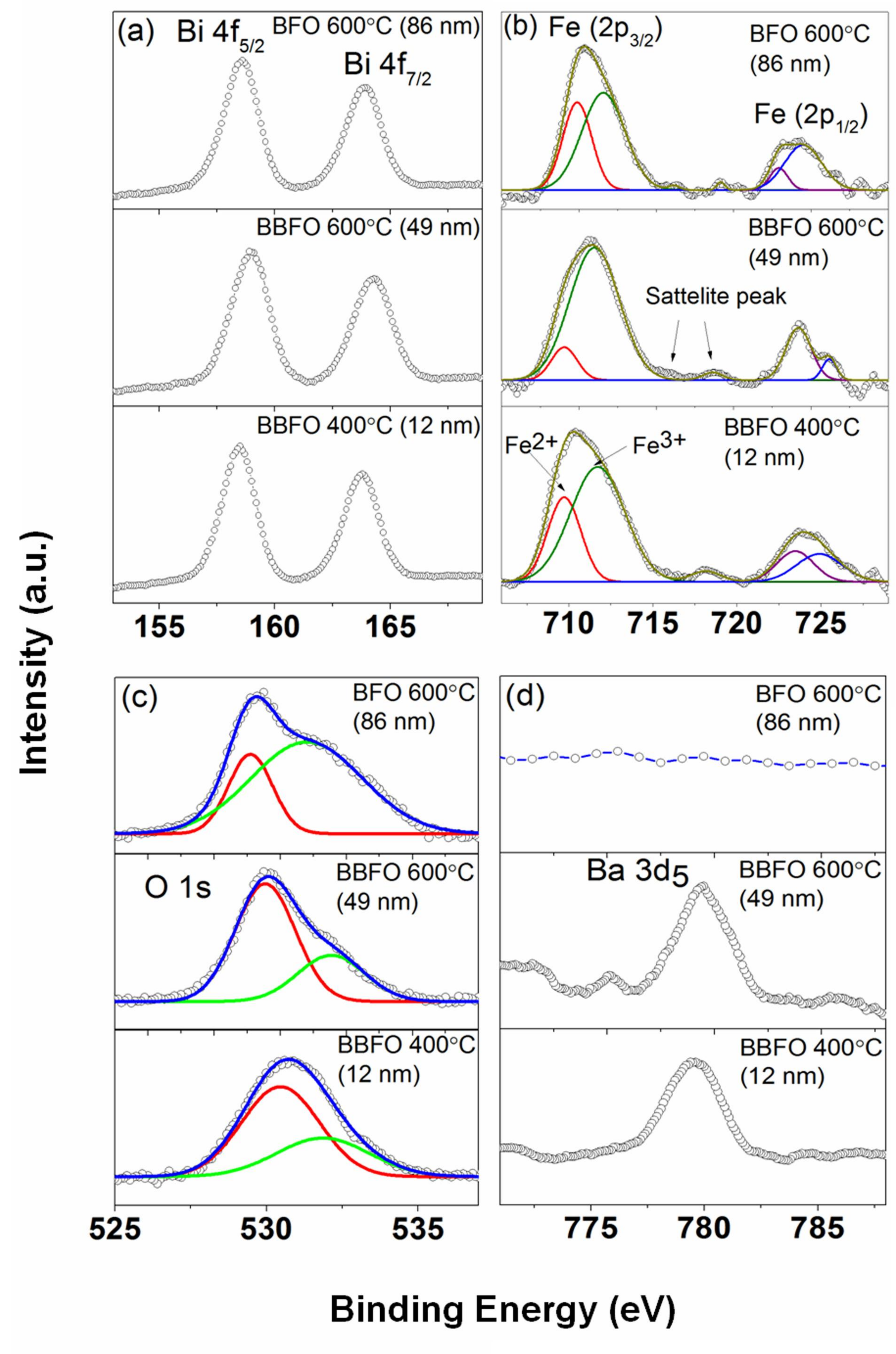}
	\caption{XPS spectra of (a) Bi 4f lines (b) Fe 2p lines (c) O 1s lines and (d) Ba 3d lines for undoped and Ba-doped BFO nanoparticles with indicated particle sizes and annealing temperature.} \label{fig2}
\end{figure}

The multiferroic properties of BFO ceramics are highly susceptible to valence states of cations and oxygen vacancies \cite{ref13}. To examine the evidence of multiple valence states of Fe and concentration of oxygen vacancies, XPS analysis has been performed for BFO powders annealed at 600$^o$C and BBFO powders annealed at 400$^o$C and 600$^o$C. The core level binding energy spectra of Bi 4f, Fe 2p, O 1s and Ba 3d orbitals are shown in figures \ref{fig2}(a)-(d) respectively. The distance between two peaks corresponding to the Bi 4f$_{5/2}$ and Bi 4f$_{7/2}$ shown in Fig. \ref{fig2}(a) is approximately 5.35 eV which suggests the existence of stable Bi$^{3+}$ ion in the doped and undoped BFO ceramics \cite{ref13}. The XPS spectra of Fe 2p shown in Fig. \ref{fig2}(b) reveals two intense peaks at around 710.5 eV and 724 eV which are attributed to the Fe 2p$_{3/2}$ and Fe 2p$_{1/2}$, respectively.The XPS spectra of Fe 2p$_{3/2}$ have been de-convoluted into two peaks at around 709.8 eV and 711.5 eV for BFO powders annealed at 600$^o$C, 709.01 eV and 711.0 eV for BBFO powders annealed at 600$^o$C, 709.7 eV and 711.6 eV for BBFO powders annealed at 400$^o$C. Generally, the binding energy peak at 709.5 eV and 711 eV are assigned to Fe$^{2+}$ and Fe$^{3+}$ respectively \cite{ref29}. The coexistence of mixed oxidation states of Fe is further confirmed by the presence of two satellite peaks at around 6 ev and 8eV above the principal peak of Fe$^{2+}$ 2p$_{3/2}$ and Fe$^{3+}$ 2p$_{3/2}$ respectively \cite{ref30}. However, a significant decrement of Fe$^{2+}$ state (by comparing the area under two de-convoluted peaks corresponding to Fe$^{2+}$ and Fe$^{3+}$ states as shown in Fig. \ref{fig2}(b)) is observed for BBFO samples compared to its undoped counterpart which can be explained by the following reaction:

\begin{align}
\mathrm{Bi^{3+} + Fe^{2+} \rightarrow Ba^{2+} + Fe^{3+}}
\end{align}


Essentially, the replacement of Bi$^{3+}$ by Ba$^{2+}$ ion is balanced by switching of Fe$^{2+}$ to Fe$^{3+}$ state. The calculated Fe$^{3+}$:Fe$^{2+}$ ratio increases from 100:61 to 100:13 by virtue of 10\% Ba-doping in BFO ceramics which has been annealed at 600$^o$C. As XPS is a surface sensitive technique the contribution of Fe$^{2+}$ state from surface impurity phases cannot be simply ruled out in case of undoped BFO ceramics. Besides, an increment of Fe$^{2+}$ state is evident for BBFO nanoparticles with reducing particle size and annealing temperature [Fig. \ref{fig2}(b)].The increased surface defects and impurity phases [Fig. \ref{fig1}(a)] with reducing particle size of BBFO nanoparticles may be attributed to the obvious increase of Fe$^{2+}$ state. 

The Fig. \ref{fig2}(c) displays the core level XPS spectra of O 1s orbital which have been de-convoluted into two peaks at around 529 eV and 531 eV. The lower binding energy peak is associated with the intrinsic O 1s core spectra and higher energy peak is ascribed to the oxygen vacant sites in BFO ceramics \cite{ref31}. It is depicted in Fig. \ref{fig2}(c) that 10 \% Ba-doping significantly curtails oxygen vacancies in BBFO nanoparticles. The highly volatile nature of Bi leads to off-stoichiometry (impurity phases) and oxygen vacancy in BFO sample. The larger Ba$^{2+}$ ion in BBFO system resist the motion of moveable Bi$^{3+}$ ion and consequently reduces impurity phases. Additionally, two Bi$^{3+}$ vacancies are balanced by creating three oxygen vacancies. Whereas, the substitution of two Bi$^{3+}$ by two Ba$^{2+}$ ions is balanced by creating one oxygen vacancy \cite{ref32}. So an overall decrease in oxygen vacancy is attained for 10 \% Ba-doped samples. 

\begin{figure}[!hh]
	\centering
	\includegraphics[width=9cm]{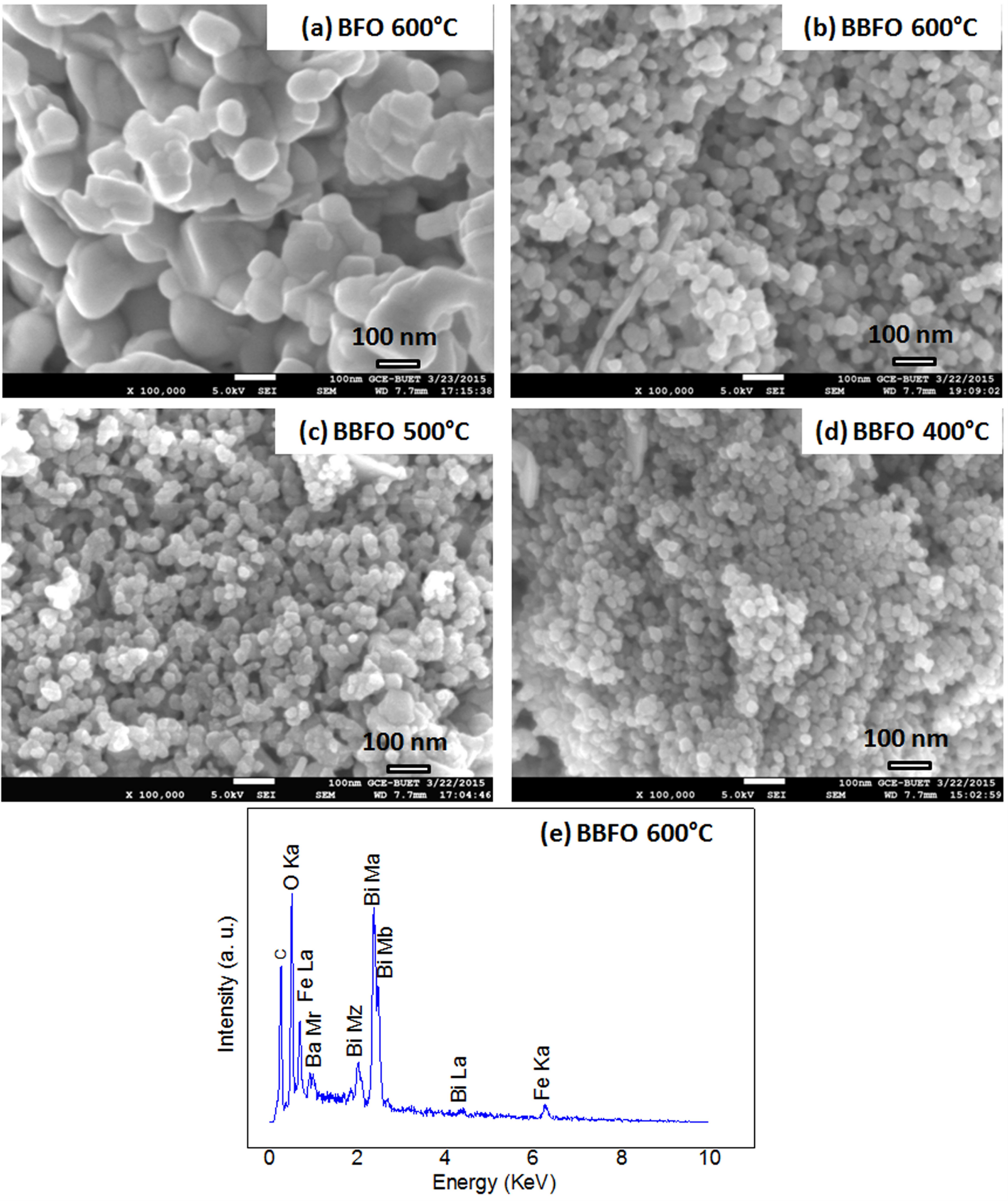}
	\caption{FESEM micrographs of (a) BFO 600$^o$C , (b) BBFO 600$^o$C , (c) BBFO 500$^o$C , (d) BBFO 400$^o$C. (e) EDX spectra recorded from BBFO 600$^o$C  FESEM image.} \label{fig3}
\end{figure}
The FESEM micrographs shown in Figs. \ref{fig3}(a)-(d) reveal the effect of Ba-doping and heat treatment temperature on the particle size of BBFO samples. The EDX spectrum [figure \ref{fig3}(e)] has been taken from figure \ref{fig3}(b) which reveals the presence of Ba in doped nanoparticles. It is apparent that the calculated average particle sizes from Debye-Scherrer equation are in good agreement with that of observed from FESEM micrographs [figures \ref{fig3}(a)-(d)] and increases with increasing annealing temperature as expected. Figs. \ref{fig3}(a) and (b) also manifest the particle size inhibition effect of Ba-doping in BFO powders. Average particle size of BBFO nanoparticles annealed at 600$^o$C is 49 nm whereas that of undoped BFO is 86 nm. The increase of particle size with annealing temperature is completely a diffusion controlled phenomenon which strongly depends on point defects like bismuth and oxygen vacancies in the lattice. As was mentioned earlier, Bi is very volatile in nature and creates many oxygen vacancies in pure BFO which enhances the diffusion mechanism and thereby plays a significant role in particle growth \cite{ref33}. The replacement of Bi ion by Ba ion scales down the concentration of lattice vacancies which may inhibit the defect induced diffusion mechanism and consequently hinder growth of BBFO nanoparticles \cite{ref16}. 


\subsection{Magnetic characterization} \label{II}
\begin{figure}[!t]
	\centering
	\includegraphics[width=8.5 cm]{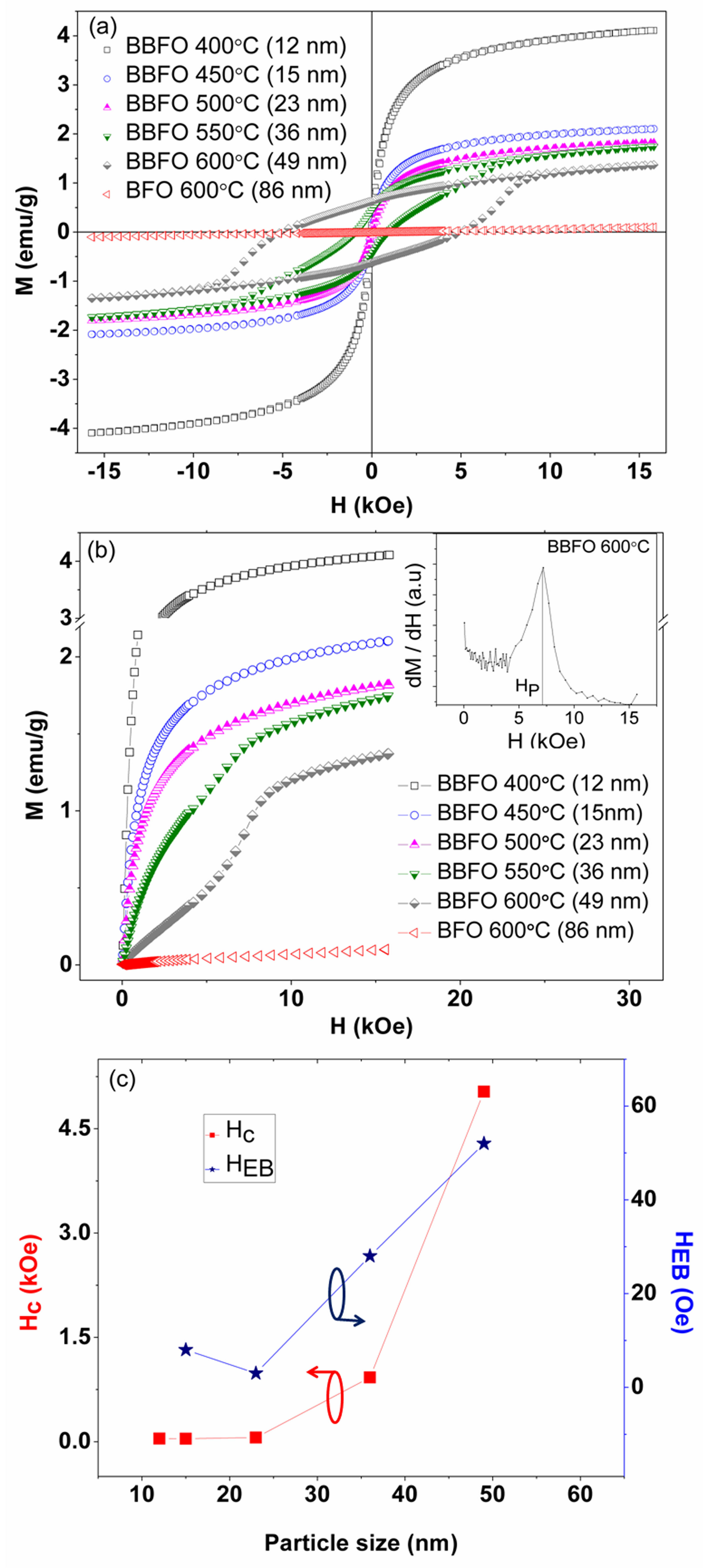}
	\caption{(a) Room temperature magnetic hysteresis loops of BFO and BBFO powders with indicated particle sizes. (b) Initial magnetization curve of BFO and BBFO powders annealed at different temperatures with inset showing derivative of the initial magnetization curve for BBFO powders annealed at 600$^o$C.(c) Variation of coercivity and exchange bias field as function of particle size.} \label{fig4}
\end{figure}
Room temperature M-H hysteresis loops of nanocrystalline BFO and BBFO powders annealed at temperatures ranging from 400$^o$C to 600$^o$C are shown in Fig. \ref{fig4}(a). The initial magnetization curves for BFO and BBFO nanoparticles annealed at different temperatures are shown in Fig. \ref{fig4}(b). The inset of the figure displays derivative of the initial magnetization curve for BBFO powders annealed at 600$^o$C. It is apparent from Fig. \ref{fig4} that undoped BFO nanoparticles annealed at 600$^o$C show nearly antiferromagnetic (AFM) nature. Whereas, BBFO nanoparticles annealed at 400-600$^o$C show nearly ferromagnetic (FM) behavior and their magnetization increases with reducing particle size. The onset of ferromagnetism in BBFO nanostructures could be attributed to the suppression of spiral spin order and structural distortion by Ba-doping which is quantified as changes in Fe-O-Fe bond angle, $\phi$ \cite{ref34, ref35, ref36}. As shown in Table \ref{Tab1}, $\phi$ increase significantly with Ba-doping which is analogous with previous investigation \cite{ref37}. The change in bond angle modifies the tilting angle of FeO$_6$ octahedron which might suppress the spiral spin structure and hence outset net magnetization in BBFO nanoparticles. On the contrary, according to Goodenough-Kanamori rule the AFM superexchange interaction between two magnetic ions with partially filled d orbital increases with increasing bond angle, $\phi$ and is strongest when  $\phi$=180$^o$ \cite{ref38, ref39, ref40}. Therefore, although AFM interaction increases due to Ba doping, the apparent ferromagnetism in Ba-doped BFO nanoparticles could be attributed to the local distortion induced FM pinning in AFM superlattice. The presence of ferromagnetic pinning in AFM structure is further substantiated by the observed coercivity, exchange bias field and metamagnetic transition in BBFO nanostructures which will be discussed progressively in this article. Furthermore, the Fe-O-Fe bond angle decreases with reducing particle size of BBFO nanoparticles [Table \ref{Tab1}]. The AFM superexchange interaction weakens with decreasing bond angle, $\phi$ which results in spin canting away from the perfect AFM structure and introduce weak FM interaction \cite{ref38, ref39, ref40}. The weakening of AFM interaction could be one of the important factors behind the apparent raise of magnetization [Fig. \ref{fig4}] with reducing particle size. Additionally, the increment in saturation magnetization with reducing particle size could also be interpreted by a particle modeled as AFM (essentially weak ferromagnetic) core and ferromagnetic surface \cite{ref9}. Essentially the Fe-O bond is disrupted at the particle surface. Uncompensated spins originate from Fe$^{3+}$ ions by means of missing oxygen ions at the particle surface. These uncompensated spins could significantly contribute to the particles overall magnetization \cite{ref9, ref10, ref11}. In BBFO nanoparticles the contribution of uncompensated spins at the particle surface increases with decreasing particle size due to the large surface to volume ratio and consequently increase saturation magnetization \cite{ref12}. Therefore, the complex interaction between these uncompensated surface spins and onset of FM interaction in BBFO nanoparticles may enhance magnetization with reducing particle size.
\begin{figure}[!hh]
	\centering
	\includegraphics[width=9cm]{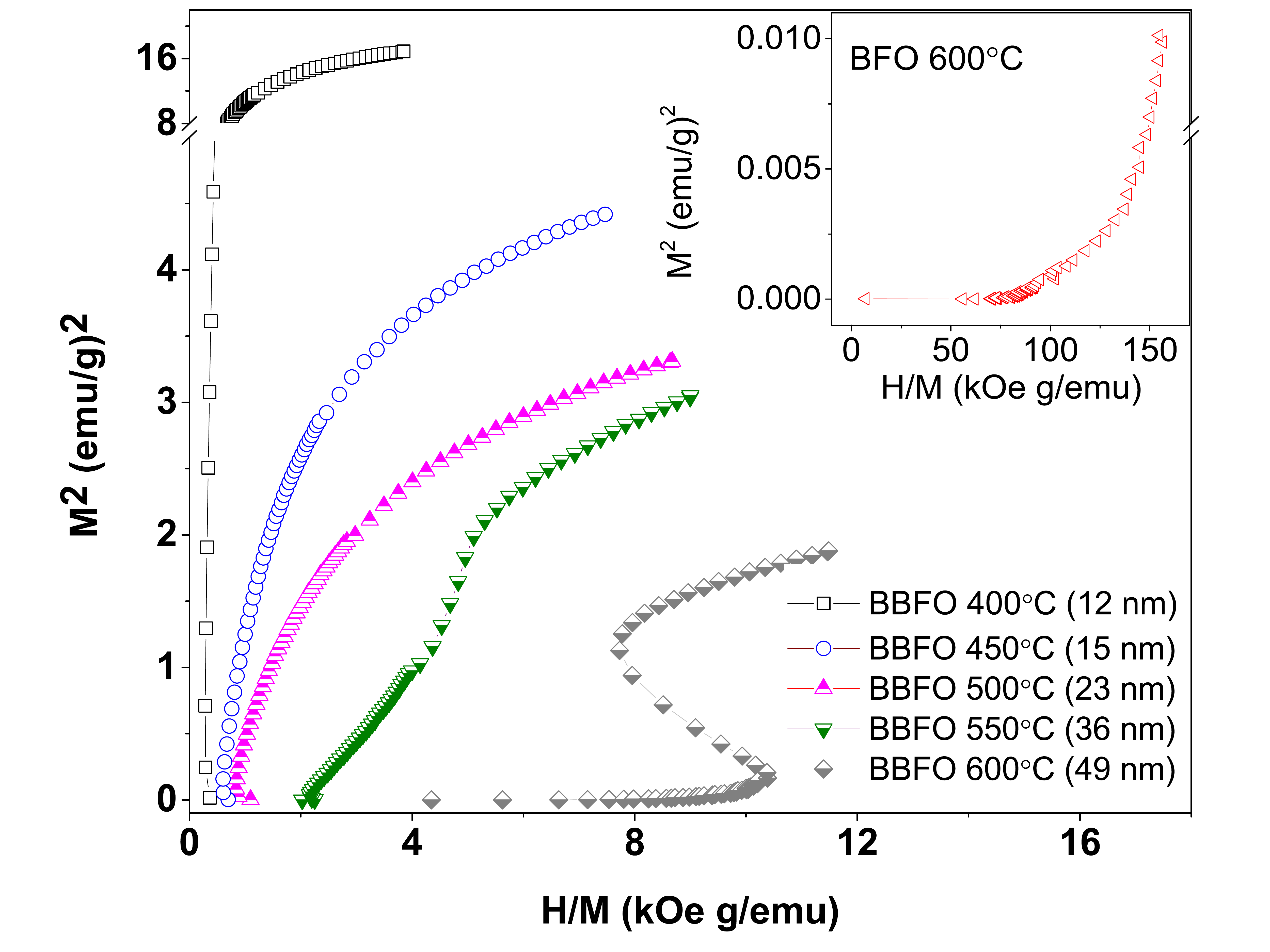}
	\caption{Arrot plots of BBFO nanoparticles annealed at 400-600$^o$C. The inset displays arrot plot of BFO annealed at 600$^o$C.} \label{fig5}
\end{figure}

From the M-H hysteresis loops the coercive field has been calculated by $H_c = (H_{c1}-H_{c2})/2$, and exchange bias (EB) field which is a measure of the shift of M-H loop, has been calculated by $H_{EB} = -(H_{c1}+H_{c2})/2$ where H$_{c1}$ and H$_{c2}$ are the left and right coercive fields at zero magnetization  \cite{ref41, ref301}. Figure \ref{fig4}(c) displays the variation of exchange bias, H$_{EB}$ and coercive field, H$_{c}$ as affected by particle size. The figure reveals that 49 nm BBFO nanoparticles show significantly high H$_{c}$ and EB field which decreases with reducing particle size. The observed H$_{c}$ value for 49 nm BBFO sample is in good agreement with some previously published values \cite{ref35,ref36}. The coercive field may originate from locking of local moments associated with local distortions in BBFO samples which could be described as ferromagnetic (FM) pinning in antiferromagnetic (AFM) super lattice \cite{ref35,ref36}. The association of pinning mechanism with the observed coercive field may be substantiated by the derivative of magnetization (dM/dH) plot as shown in the inset of Fig. \ref{fig4}(b). The pinning field, H$_{p}$ which is obtained from the derivative plot is in comparable range with the observed H$_{c}$ value indicating the presence of spin pinning mechanism in BBFO nanoparticles \cite{ref42}. Moreover, the noticed EB field for 49 nm and 36 nm BBFO samples also clearly demonstrates the presence of complex interaction between different magnetic phases which could be either FM pinning in AFM core or uncompensated spin induced FM surface and AFM core or both \cite{ref43,ref44}. Furthermore, with decreasing particle size the AFM superexchange interaction become weaker which results in weak AFM-FM interaction. The outset dominance of uncompensated surface spins followed by waning of AFM-FM interaction and easy flip nature of those surface spins may result in decreasing trend of H$_{c}$ with decreasing particle size [Fig. \ref{fig4}(c)]. Notably, in Ref. \cite{ref15}, the coercivity was increased with reduced particle size which were varied due to compositional variation of Bi$_{1-x}$Ba$_{x}$FeO$_3$ (x = 0.05-0.30) nanocrystalline system.
	

In Fig. \ref{fig4}(b) an upturn of magnetization is evidenced for BBFO nanoparticles annealed at 550$^o$C and 600$^o$C with applied fields in the range of 5 to 8 kOe which is attributed to the magnetic field induced AFM to FM metamagnetic transition \cite{ref21} in this nanoparticle system. This behavior also demonstrates the presence of mixed magnetic phases \cite{ref21}. Applied field at and above critical pinning field, H$_{p}$ the pinned ferromagnetic regions in the antiferromagnetic matrix of the particles begin to align and grow rapidly and thereby metamagnetic transition comes into view [Fig. \ref{fig4}(b)]. An indication for the presence of mixed magnetic phases and interaction between them are justified by the observed H$_{EB}$ values \cite{ref7,ref41}. The diminishing EB field in response to decreasing particle size as shown in the figure \ref{fig4}(c) indicates the waning of AFM-FM interaction which also justifies the corresponding disappearance of metamagnetic transition in BBFO nanoparticles with reducing particle size.

To investigate the order of magnetic transition, Arrott-plots (M$^2$ vesrsus H/M) shown in Fig. \ref{fig5} have been constructed for both BFO and BBFO nanoparticles. According to Banerjee criterion \cite{ref45} the negative slope in Arrott-plot for 49 nm BBFO nanoparticles indicates the first order metamagnetic transition while 36 nm BBFO nanoparticles show second order metamagnetic transition. With further reduction in particle size metamagnetic transition disappears with the manifestation of normal FM behavior as evidenced in initial magnetization curves [Fig. \ref{fig4}(b)] and their Arrott-plots [figure \ref{fig5}]. Notably, in previous investigations pure BFO also shows second order metamagnetic transition at very high field, 20 T and low temperature 5 K, which has been attributed to the destruction of AFM spiral spin order \cite{ref46,ref47}. The observed room temperature metamagnetic transition at low applied field in case of BBFO nanoparticles [figure \ref{fig5}] may be due to the modification of long range AFM spin order as a result of Ba-doping which requires further investigation. In the next stage of this investigation, we have carried out electrical characterizations of the synthesized samples. 

\subsection{Ferroelectric characterization} \label{III}
\begin{figure}[!hh]
	\centering
	\includegraphics[width=9cm]{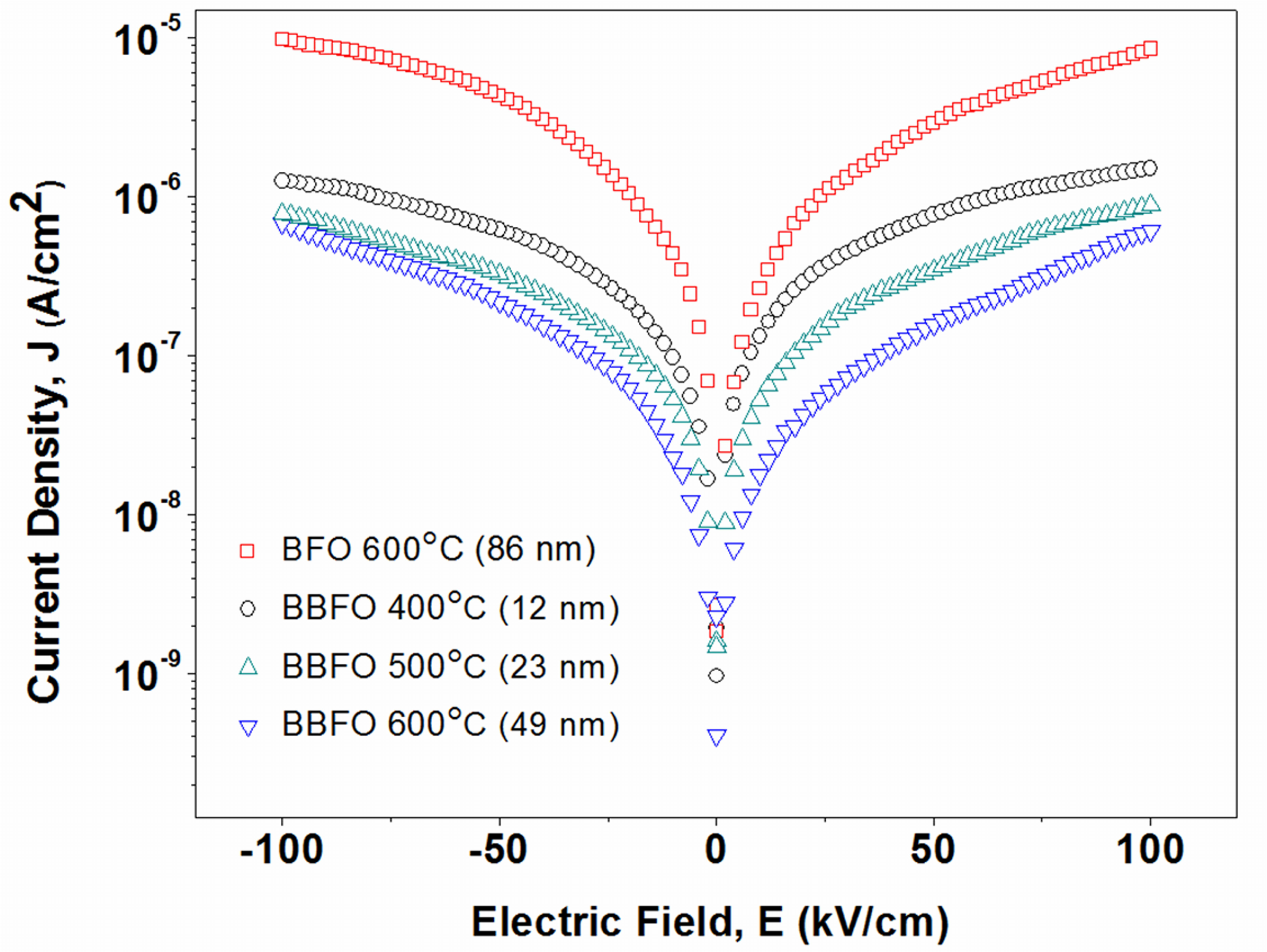}
	\caption{Room temperature leakage current density, J versus applied electric field, E plot for BFO and BBFO powders annealed at different temperatures.} \label{fig6}
\end{figure}

First of all, to examine the leaky behavior of BFO and BBFO samples leakage current density, J versus electric field, E measurements were performed for an applied field of up to 100 kV/cm. Figure \ref{fig6} reveals the effect of Ba-doping and particle size on leakage current density of BBFO nanoparticles. The high leakage current of undoped BFO in this study is predominantly connected with impurity phases, oxygen vacancies and electron hopping from Fe$^{2+}$ to Fe$^{3+}$ state \cite{ref32,ref48}. It is worth mentioning that 10\% Ba-doping in BFO significantly diminishes impurity phases, oxygen vacancies and Fe$^{2+}$ state which have been confirmed by XRD and XPS analysis. The replacement of smaller Bi$^{3+}$ by larger Ba$^{2+}$ ion is supposed to hinder ion mobility and thereby may also subsides leakage current \cite{ref49}. Moreover, it is clear from figure \ref{fig6} that leakage current density of BBFO nanoparticles increases with reducing particle size and annealing temperature. Some previous studies showed that electrical conductivity of oxide nanoparticles \cite{ref50,ref51,ref52} increases with decreasing particle size. It is speculated that the donor effect \cite{ref52} of grain boundaries and easy path for ionic diffusion \cite{ref50} may result in increased electrical conductivity in fine particles. The iron rich Bi$_2$Fe$_4$O$_9$ impurity phase mostly occupies at the grain boundaries \cite{ref8,ref23} which is also perceived to accelerate the conduction mechanism in fine particles. Although the quantification of small impurity phases in the BBFO samples is very difficult from XRD results, the possibility of increasing these phases as well as leakage current with decreasing annealing temperature cannot be simply ruled out. 
\begin{figure}[!hh]
	\centering
	\includegraphics[width=9cm]{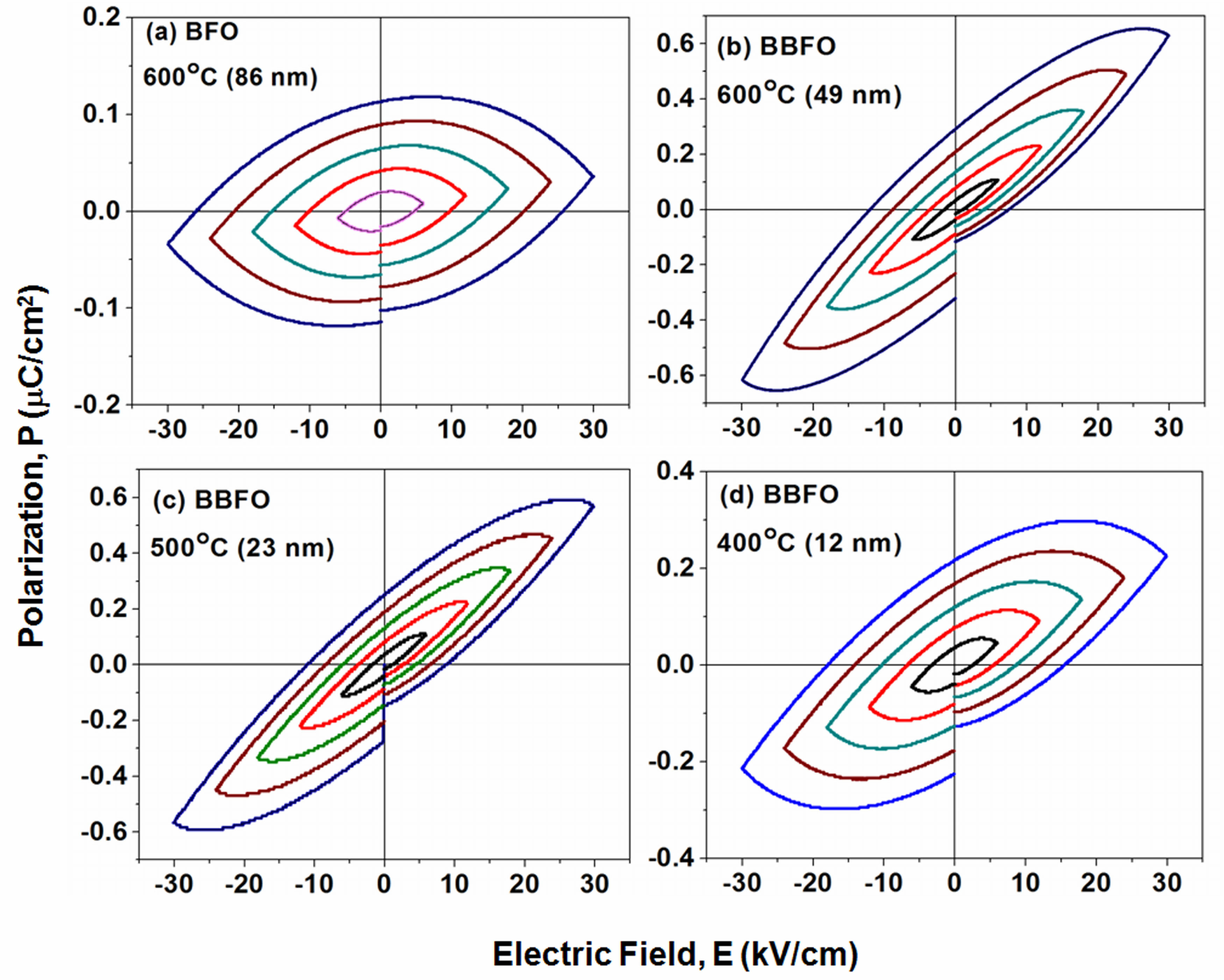}
	\caption{Room temperature ferroelectric hysteresis loops of BFO powders annealed at (a) 600$^o$C and BBFO powders annealed at (b) 600$^o$C, (c) 500$^o$C, (d) 400$^o$C.} \label{fig7}
\end{figure}
We have also carried out ferroelectric measurement to establish polarization versus electric field hysteresis loops (P-E) at applied field with a maximum value of $\pm$ 30 kV/cm for BFO and BBFO nanoparticles. The P-E loops of nanoparticles annealed at different temperatures are shown in Fig. \ref{fig7} which actually demonstrates the effect of Ba-doping and particle size reduction. Undoped BFO exhibits a round shaped P–E loop as a result of its high leakage current [Fig. \ref{fig6}]. Whereas, significantly improved ferroelectric polarization has been observed for 10\% Ba-doped samples owing to reduced leakage current demonstrated in Fig. \ref{fig6}. Previous investigations confirmed that cation substitution enhances ferroelectric property of BFO by virtue of reduced leakage current \cite{ref34,ref53,ref54}. Moreover, a particle size dependent polarization for BBFO nanoparticles is vivid in Fig. \ref{fig7}. A pioneering work \cite{ref28} done in 2007 showed that polarization of BFO diminishes with reducing particle size. Essentially, the rhombohedral distortion originates a displacement of Bi$^{3+}$ (with lone pair 6s electron) ion from its position regarding to its ideal cubic perovskite structure which results in A-site polarization in BFO \cite{ref3}. However, with decreasing particle size the rhombohedral symmetry of BFO relaxes toward cubic symmetry which is accompanied by an increment of lattice volume and decrease of c/a ratio  \cite{ref28}. The downturn of c/a ratio along with an upturn of lattice volume with reduced particle size as depicted in Table \ref{Tab1} may be one of the possible reasons for the observed size dependent polarization in this study. Particle size dependent ferroelectric polarization has also been reported for BaTiO$_3$ nanoparticles previously \cite{ref55}. Below 70 nm BaTiO$_3$ nanocrystals relaxes its tetragonality towards cubic symmetry and became paraelectric gradually. Moreover, the increased leakage current with diminishing particle size [ Fig. \ref{fig6}] might be another salient feature for the decreased polarization values.

\section{Conclusions}
In summary, size dependent magnetic and electrical properties for 10\% Ba-doped BiFeO$_3$ nanoparticles have been explored. A threefold enhancement in saturation magnetization was observed in $\sim$ 12 nm Bi$_{0.9}$Ba$_{0.1}$FeO$_3$ nanoparticles due to the combined effect of cation substitution and size confinement. The coercivity of the synthesized nanoparticles was also found to decrease significantly with reducing particle size and below 23 nm particle size the coercivity is almost negligible.  This actually indicates their soft nature and potentiality in device applications where a negligible coercivity at room temperature is crucially effective. The Bi$_{0.9}$Ba$_{0.1}$FeO$_3$ nanoparticles with average particle size 49 nm and 36 nm show metamagnetic transition which has been found to disappear with diminishing particle size. Additionally, a reduction in moveable charges owing to Ba-doping is attained which has been confirmed by XPS analysis, leakage current and ferroelectric measurements. The tunable multiferroic properties by controlling particle sizes of Bi$_{0.9}$Ba$_{0.1}$FeO$_3$ nanoparticles is auspicious and may open a door to think about its processing (annealing) temperature and average grain size intending its applications in miniaturized devices. Moreover, the room temperature first order metamagnetic transition at low applied field in Bi$_{0.9}$Ba$_{0.1}$FeO$_3$ nanoparticles annealed at 600$^o$C may find great interest in the subsequent investigations due to its probable magnetocaloric effect \cite{ref19,ref20,ref21}. 

\section{Acknowledgements}
The authors would like to thank Ministry of education, the Peoples Republic of Bangladesh for its financial support.

\end{document}